\documentclass[aps,prl,twocolumn,superscriptaddress,showpacs]{revtex4-1}
\usepackage{amsmath,amssymb,amstext,graphicx,bm}
\usepackage{graphicx,color,bm}
\usepackage[english]{babel}
\usepackage{dsfont}

\usepackage{epsfig}
\begin{document}

\title{Giant edge spin accumulation in a symmetric quantum well with two subbands}

\author{Alexander Khaetskii}
\affiliation{Department of Physics, University at Buffalo, SUNY, Buffalo, NY 14260-1500}

\author{J. Carlos Egues}
\affiliation{Instituto de F\'{\i}sica de S\~ao Carlos, Universidade de S\~ao Paulo, 13560-970, S\~ao Carlos, S\~ao Paulo, Brazil}

\date{\today}

\begin{abstract}
We have studied  the edge spin accumulation due to an electric current in  a high mobility two-dimensional electron gas 
formed in a symmetric well with two subbands.  This study is strongly motivated by the recent experiment of Hernandez {\it et al.} [Phys. Rev. B {\bf 88}, 161305(R) (2013)] who demonstrated the spin accumulation near the edges of a  symmetric bilayer GaAs structure in contrast to no effect in  a single-layer configuration. 
The intrinsic mechanism of the spin-orbit (SO) interaction  we consider arises from the coupling between two subband states of opposite parities. 
We obtain a parametrically large magnitude of the edge spin density for a two-subband well as compared to the usual single-subband structure. We show that the presence of a gap in the system, i.e., the energy separation $\Delta$ between the two subband bottoms, changes drastically the picture of the edge spin accumulation.  The gap value governs the effective strength of the inter-subband SO interaction which provides a controllable crossover from
the regime of weak spin accumulation to the regime of strong one by varying  the Fermi energy (electron density) and/or $\Delta$. 
We estimate that by changing the gap $\Delta$ from zero up to $1\div 2$ K, the magnitude of the effect changes by three orders of magnitude.
This opens up the possibility for the design of new spintronic devices. 
\end{abstract} 


\maketitle

       Spin current and spin accumulation \cite{Engel,We} which appear due to the spin-orbit (SO)  coupling in the presence of 
electric currents are topics of great current interest which are  important for the future of spin electronics \cite{Zutic}.  There are two distinct SO  mechanisms,  the extrinsic one due to the Mott asymmetry in the electron scattering off  impurities \cite{Dyak,Hirsh,Khaet1,Khaet2}, and the intrinsic one \cite{Murakami,Sinova} due to  SO induced splitting of the electron spectrum.  The edge spin-density accumulation, related to either the Mott asymmetry by impurities \cite{Kato} (2D electrons) or the intrinsic mechanism (2D holes) \cite{Wunder,Nomura}, has been experimentally  observed.  
\par
     It is known [\onlinecite{We,Tserk,Bleib,Adagideli}] that in the diffusive regime (and when the spin diffusion length is much larger than the mean free path) the edge spin density is entirely due to the spin flux coming from the bulk.  In contrast, the physics of the edge spin-density accumulation for the intrinsic mechanism  in the opposite case of strong SO splitting \cite{Nikolic} only recently has been understood [\onlinecite{Usaj,Zyuzin,Silvestrov,Khaet,Khaetskii}]. This includes  the experimentally important case of a diffusive sample with a large SO splitting of the spectrum so that the spin-precession length is smaller than the mean free path. This  we term the quasi-ballistic regime [\onlinecite{Khaet},\onlinecite{Khaetskii}].  In particular, for 2D holes in this regime the edge spin-density, which is  due to the spin current from the bulk, is parametrically smaller than the density generated upon the boundary scattering \cite{Khaetskii}. 
\par
Recently,  using Kerr rotation spectroscopy, 
 Hernandez {\it et al.} [\onlinecite{Hernandez}] demonstrated electric-current induced spin accumulation near the edges of a high-mobility two-dimensional electron gas in a {\it symmetric} bilayer GaAs structure in contrast to {\it no accumulation} in a single-layer configuration  \cite{note2}. This result is interesting and intriguing in many aspects. 
The observed effect is quite large
despite the fact that the electric field in the high-mobility channel is $300 \div 400$  times smaller than that in the experiment  by Kato {\it et al.} [\onlinecite{Kato}],  where for a GaAs sample the result was explained by the extrinsic interaction with impurities. 
Note that the structure studied in [\onlinecite{Hernandez}] has inversion symmetry and therefore the usual Rashba term \cite{Vasko} is absent.  On the other hand, the linear-in-momentum term \cite{Dyak1} originating from a cubic Dresselhaus term is known not to lead  to a spin current in the bulk. 
A significant difference between the observed edge spin density in the two-subband vs. the one-subband cases has motivated us to look for the explanation of this phenomenon using the inter-subband Rashba-like Hamiltonian arising in two-subband wells~\cite{Bernardes},\cite{casalverini}. 
\begin{figure}
\begin{center}
\centerline{\resizebox{2.915in}{!}{\includegraphics{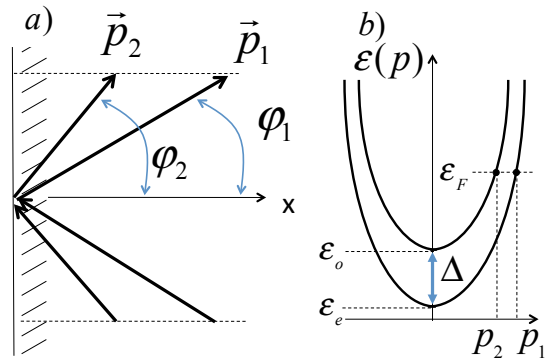}}}
\end{center}
\caption{Schematics of the boundary specular scattering in the presence of SO coupling. Plus and minus modes are shown for the same energy and the same wave vectors along the boundary.} 
\label{fig:Spin1}
\end{figure}

Here we follow the method proposed in \cite{Khaetskii} to calculate the edge spin density which appears due to boundary scattering [Fig.~1(a)] in the quasi-ballistic regime for a Rashba-like Hamiltonian \cite{Bernardes},\cite{casalverini} describing the two-subband well [Fig.~1(b)]. 
In this quasi-ballistic regime the characteristic length of the spin accumulation near the boundary is smaller than the mean free path. Since the latter is around 30 $\mu m $~\cite{Hernandez}, it indeed exceeds all the characteristic lengths of our theory. 
 We have explained the experimental results, in particular,  the large magnitude of the edge spin density for the two-subband sample compared to the usual  single-band structure with either the Rashba or Dresselhaus interactions. 
\par
{\it Two bands vs one band case.}
Interestingly, we have found that despite the problem in question resembling very much the usual Rashba problem (there are two copies of them because each state is doubly degenerate), the presence of the gap $\Delta$ between two sub-band edges [Fig.~1(b)]  changes the physics of the edge spin  accumulation completely. This happens because the gap magnitude governs the effective strength of the inter-subband SO interaction leading to different solutions compared to the one-band Rashba case for the occupation numbers of the incoming states participating in the boundary scattering, [Fig.~1(a)]. 
\par
The physics is now determined by the value of the parameter $\xi=2\eta k_F/ \Delta \equiv L_{\Delta}/L_s$ [Fig.~2a],  where $L_{\Delta}=\hbar v_F/\Delta$, $L_s=\hbar^2/2m\eta$ are the coherence  and the  spin-precession lengths.  Here $p_F=\hbar k_F=m v_F$ is the Fermi momentum, $\eta$ is the inter-subband SO coupling constant \cite{Bernardes,casalverini}.  The  parameter $\xi$ can have an arbitrary value even for a small $\eta$ since the gap $\Delta$ can be made much smaller than the Fermi energy. 

\par
First of all, the presence of the gap changes the "helicity"  direction which corresponds to the eigenvectors of the Hamiltonian (\ref{H1}), see Eqs.~(S2), (S3) in  the Supplemental Material (SM).  
For the two sub-band problem the helicity axis for a given momentum $p$ is defined by the angle $\theta(p)$ with the z-axis (the normal to the 2D plane),  $\cos \theta =1/\sqrt{1+ (2\eta p/\hbar \Delta)^2}$.  When $\Delta$ is much bigger than the SO energy $2\eta p/\hbar$, then the helicity axis coincides with the z-axis. In the opposite case when  $\Delta \rightarrow 0$ and  the SO energy dominates, the helicity axis lies within the $x-y$ plane, as it should be for usual Rashba model (which corresponds to $\theta \rightarrow \pi/2$). Therefore, $\sin \theta = (2\eta p/\hbar)/\sqrt{(2\eta p/\hbar)^2+ \Delta^2}$ determines the {\it effective} strength of SO coupling, and the usual Rashba single-band model (in our case we have two copies of them) corresponds to   a limit of strong SO coupling ($\sin \theta =1$). 
\par 
As it has been shown in \cite{Khaet,Khaetskii} [see also Eq.~(\ref{spin2}) below], because of the unitarity of the boundary scattering the magnitude of the edge spin accumulation is proportional to the difference  $f_1(\varepsilon_F,k_y) -f_2(\varepsilon_F,k_y) $  between the distribution functions of the incoming electron states belonging to  the sub-bands 1 and 2 for a given Fermi energy and given wave vector $k_y$  along the boundary, see Fig.~1. ( In the case of one-band Rashba model the incoming states belong to the branches of opposite helicities).  These distribution functions are found from the solution of the kinetic equation for the spin-density matrix in the bulk (2D) of the sample in the presence of impurity scattering and an electric field \cite{Khaetskii1}, \cite{Shytov} and used as the input parameters for the boundary scattering problem. 
\par
The important point is that in the one-band Rashba case the difference of the distribution functions in question is of the third order  with respect to $p_1-p_2=\hbar/L_s$ for any reasonably short-ranged impurity potential in the bulk (when the correlation radius of the impurity potential $d$ is much smaller than $L_s$).  This has been rigorously proven in  Ref.~\cite{Khaetskii} (see Eqs.~(14-16) in \cite{Khaetskii}).  
Thus, in the leading order (i.e. $\propto 1/L_s$)
 the above mentioned distribution functions take the form which
they have in the absence of any spin-orbit coupling. 
The first order effect appears only for very smooth impurity potential when $d > L_s$, and the magnitude of the edge spin density for the  one-band Rashba case is given by the expressions \cite{Khaetskii}
\begin{equation}
\langle S_z \rangle \simeq \frac{k_E}{L_s}\frac{d^2}{L_s^2} \,\,\, at \,\, d \ll L_s; \,\, \langle S_z \rangle \simeq \frac{k_E}{L_s} \,\,\, at \,\, d>L_s. 
\label{one_band}
\end{equation}
Here $k_E=eE\tau_{tr}/\hbar$,  $e$ is the modulus of the electron charge, $E$ is the magnitude of the in-plane (driving) electric field directed along the y-axis and $\tau_{tr}$ is the transport scattering time due to the impurities in the bulk of a sample.  
Note that we consider in our paper the case $d\ll L_s$, which is only the realistic one.
\par Note that the cancellation of the effect for the Rashba model in the leading order (i.e. $\propto 1/L_s$)  happens when the transition rates between branches of opposite helicities have the same strength as the transition rates within the branch of the same helicity (one of the necessary conditions). That is why for smooth impurity potential ($d>L_s$) which cannot support the transitions  between branches of opposite helicities,  one observes the recovering of the first order effect, see Eq.~(\ref{one_band}). 
\par
From the above considerations we can immediately understand the role of the gap $\Delta$ in the two-subband model considered in this work. 
As explained above, the effective SO interaction decreases with increasing $\Delta$. This causes suppression of the inter-subband transition rates, since these transitions are accompanied by spin flip,  as compared to the intra-subband ones, which do not need spin flip. The suppression factor is $\sin^2 \theta $, which is the probability of spin-flip (see also SM, Sec. III). Since the inter-band  and intra-band rates are different now, this prevents the complete cancellation that occurs for the quantity $f_1(\varepsilon_F,k_y) -f_2(\varepsilon_F,k_y) $ in the one-band Rashba model, and leads to the recovery of the first order effect with respect to small splitting $p_1-p_2\ll p_F$ even for a short-ranged ($d\ll L_s$) impurity potential. 
\par 
As we discuss later on, for a GaAs structure similar to that used in Ref.  [\onlinecite{Hernandez}], it is enough to change the gap $\Delta$ from zero up to about $1\div 2$ K in order  to increase the magnitude of the effect by the three orders ("giant effect"). The most pronounced change happens at $L_{\Delta} \simeq 0.5 L_s$,  where the edge spin density $S_z(x)$ is maximized and its order of magnitude is given by $k_E/L_s$,  which is parametrically larger than in the single-band case [Fig.~2a]. 
\par
We consider specular scattering [i.e., a straight boundary for simplicity, Fig.~1(a)] and a Fermi energy much larger than the gap $\Delta$ between the subbands, i.e., $\varepsilon_F \gg \Delta$ \cite{EF}. 
Moreover, the SO interaction is weak ($\eta k_F \ll \varepsilon_F$) and therefore "coherent''  and spin-precession length scales are large compared to the Fermi wave length, $L_{\Delta}, L_s \gg  \lambda_F=2\pi/k_F$. 
The ratio $L_{\Delta}/ L_s$ can be arbitrary. 
Our calculation shows that the characteristic spatial scale of the edge spin density is $ \Lambda=L_{\Delta} L_s/\sqrt{L_{\Delta}^2+ L_s^2}$.  

\paragraph*{Model Hamiltonian. } The Hamiltonian of a symmetric quantum well with two subbands  and inter-subband-induced SO 
interaction resembles that of the ordinary Rashba
model.  In contrast to the latter, the intersubband SO interaction is nonzero
even in symmetric structures with the $4 \times 4 $ Hamiltonian is \cite{Bernardes}, \cite{casalverini}
\begin{eqnarray}\label{H1}
H &=& (\frac{p^2}{2m} +\varepsilon_{+} ) 1 \otimes 1 -\varepsilon_- \tau_z \otimes 1
+(\frac{\eta}{\hbar} ) \tau_x \otimes  (p_x \sigma_y-p_y \sigma_x). 
\end{eqnarray}
Here $\otimes$ means a direct tensor product,  $m$ is the effective mass, $\varepsilon_{\pm}=(\varepsilon_o\pm \varepsilon_e)/2$, 
$\varepsilon_e$ and $\varepsilon_o$ are quantized energies of the lowest (even) and first
excited (odd) subbands, respectively, measured from the bottom of the
quantum well, $\tau_{x,y,z}$  denote the Pauli matrices describing
the subband (or pseudospin) degree of freedom, and $\sigma_{x,y,z}$
are Pauli matrices referring to the electron spin. The inter-subband SO coupling $\eta$ (which has the dimensionality of square of charge) is expressed [\onlinecite{Bernardes}] in terms of the gradients of the 
Hartree-type contribution to the electron potential, the external gate and doping potentials, and  the structural quantum-well potential profile. 
Note that the gap is $\Delta=\varepsilon_o- \varepsilon_e=2\varepsilon_-$. 
\par
{\it Theoretical approach.} To calculate the edge spin density in the quasi-ballistic regime we follow the method developed in Refs.~\cite{Khaetskii,Khaet} for the case of the single-subband Rashba Hamiltonian. Assuming that the spatial scale of the edge spin accumulation $ \Lambda$ is much smaller than the mean free path $l$, we solve the edge spin problem by the method of scattering states, i.e., we find the exact quantum mechanical solution of the electron scattering by an impenetrable straight boundary [Fig.~1(a)] at a given Fermi energy.  These solutions are then used in the calculation of the (mean) spin density profile.  
The populations of the incoming states are found from the solution of the kinetic equation for the spin-density matrix in the bulk (2D) of the sample in
 the presence of electric field, see SM. 
\par
The Hamiltonian (\ref{H1}) has 4 eigensolutions $\Psi_{i,s}$,  $\Psi_{1,\uparrow},\,\, \Psi_{1,\downarrow}, \,\, \Psi_{2,\uparrow},\,\, \Psi_{2,\downarrow}$ (see SM for their explicit form) with the corresponding energy spectrum 
\begin{equation}
\varepsilon_{1,2}(p)= \frac{p^2}{2m}+\varepsilon_+ \mp \sqrt{\varepsilon_-^2 +\eta^2 p^2/\hbar^2}, 
\label{spect}
\end{equation}
where the subscript $i=1, 2$ corresponds to the lower (higher) in energy sub-band. Each sub-band is doubly degenerate with respect to the "spin direction'' $ s=\uparrow, \downarrow$ (Kramers pairs). 
Upon  scattering by the straight boundary where energy and momentum $p_y$ along the boundary are conserved, the states in the pair $\Psi_{1,\uparrow}(\varphi_1,\theta_1),\Psi_{2,\downarrow}(\varphi_2,\theta_2)$ mix up 
and form two  {\it scattering}  states, Eqs. (\ref{1up}),(\ref{2down})
 [similarly for the pair $\Psi_{1,\downarrow}(\varphi_1,\theta_1), \Psi_{2,\uparrow}(\varphi_2,\theta_2)$]. 
For this pair of scattering states, we have the following boundary condition for the scattering by a hard wall located at $x=0$ [Fig.~1(a)]
\widetext
\begin{eqnarray}
\tilde{\Psi}_{1,\uparrow}(x,y)|_{x=0}=e^{ik_yy}[\Psi_{1,\uparrow}(\pi -\varphi_1,\theta_1)e^{-ik_1x}+ F_{1,\uparrow}^{1,\uparrow}\Psi_{1,\uparrow}(\varphi_1,\theta_1)e^{ik_1x}+     
          F_{1,\uparrow}^{2,\downarrow}\Psi_{2,\downarrow}(\varphi_2,\theta_2)e^{ik_2x}]|_{x=0}=0, 
\label{1up} \\
\tilde{\Psi}_{2,\downarrow}(x,y)|_{x=0}=e^{ik_yy}[\Psi_{2,\downarrow}(\pi -\varphi_2,\theta_2)e^{-ik_2x}+ F_{2,\downarrow}^{1,\uparrow}\Psi_{1,\uparrow}(\varphi_1,\theta_1)e^{ik_1x}+     
          F_{2,\downarrow}^{2,\downarrow}\Psi_{2,\downarrow}(\varphi_2,\theta_2)e^{ik_2x}]|_{x=0}=0,
\label{2down}
\end{eqnarray}
\endwidetext 
\noindent with $p_1^2=\hbar^2(k_y^2+k_1^2)$, $p_2^2=\hbar^2(k_y^2+k_2^2)$, $\varepsilon_{1}(p_1)=\varepsilon_{2}(p_2)=\varepsilon$. 
The momenta $p_1$, $p_2$ describe states belonging to subbands 1 and 2 for a given energy $\varepsilon$, see Fig.~1(b).  
The angles $\varphi_1$, $\varphi_2$ (between the corresponding momenta and the positive direction of the $x$-axis) are expressed as $\sin (\varphi_1)=\hbar k_y/p_1$ and $\sin (\varphi_2)= \hbar k_y/p_2$. The angles $\theta_1, \theta_2$ are defined via $\cos \theta_{1,2}=1/\sqrt{1+(2\eta p_{1,2}/\hbar\Delta)^2}$.  The expressions for the scattering amplitudes ( $F_{1,\uparrow}^{1,\uparrow}$, etc.) and the corresponding components of the unitary scattering matrix $\hat{S}$ are presented in SM.  Similar equations can be written for the pair $\Psi_{1,\downarrow}(\varphi_1), \Psi_{2,\uparrow}(\varphi_2)$, and the corresponding scattering matrix elements  are also determined. 
\par
The expectation value of the z component of the spin as a function of coordinates is given by the following expression:
\begin{eqnarray}
\langle S_z(x) \rangle = \sum_{i,s}\int \frac{dk_y}{(2\pi)^2}  \frac{d\varepsilon}{v_{x,i}}f_i(\varepsilon,k_y) \nonumber \\
\times \langle \tilde{\Psi}_{i,s}(x)|\hat{S}_z|\tilde{\Psi}_{i,s}(x) \rangle 
\label{average}
\end{eqnarray}
Here  $f_i(\varepsilon,k_y)$ is the distribution function of the electron state in the sub-band $i$ for a given energy and given wave vector $k_y$  along the boundary and the group velocity is $v_{x,i}=\partial \varepsilon_i/\partial p_x$. 

We can then calculate  the most important part of the edge spin density which is smooth on the scale of the Fermi wave length \cite{note1} and  involves the interference of the outgoing waves [two last terms in Eqs.\ (\ref{1up}) and (\ref{2down})]. The corresponding formula for $ \langle S_z(x) \rangle$ valid for general values of the parameters (in the case when both subbands are occupied) is  presented in the SM.  In the most important case $p_1-p_2 \ll p_F$,  when the energy separation between two sub-bands  $\sqrt{\Delta^2+4\eta^2k_F^2}$ is much smaller than the Fermi energy, 
 the edge spin density for arbitrary values of the parameter $\xi=L_{\Delta}/L_s$ reads
%\widetext
\begin{eqnarray}
\langle S_z(x) \rangle = - \sin^2\theta \int \frac{dk_y k_y}{(2\pi)^2} \frac{d\varepsilon }{\varepsilon_F}  
\sin\left (\frac{x}{\Lambda\sqrt{1-(k_y/k_F)^2}}\right)  \nonumber \\
\times [ f_1(\varepsilon,k_y)- f_2(\varepsilon,k_y)]. 
%\nonumber \\
    \label{spin2}
\end{eqnarray}
%\endwidetext
Here $\varepsilon_F=p_F^2/2m$ is the Fermi energy. While deriving Eq.~(\ref{spin2}),  we used  that $\theta_1-\theta_2 \ll \theta_{1,2}$, and  $\varphi_2 -\varphi_1 \ll \varphi_{1,2}$. 
The difference of the distribution functions entering  Eq.~(\ref{spin2}) is calculated in the SM assuming the set of inequalities $k_F^{-1}\ll d\ll L_s$, where $d$ is the correlation radius of the impurity potential in the bulk of the structure.  The first condition means that the scattering in the bulk  is of the small-angle type. Both conditions are fulfilled for a high mobility GaAs structure. The final result derived from Eq. (\ref{spin2}) reads
\begin{equation}
\langle S_z(x) \rangle = \frac{3 k_E }{L_s} \Phi(\xi) J(x/\Lambda); \,\,\, \Phi(\xi)=\frac{\xi}{(2\xi^2+1)\sqrt{\xi^2+1}}. 
\label{result}
\end{equation}
with the spatial dependence  given by the integral
\begin{equation}
J(\frac{x}{\Lambda})=\int_{0}^{1}\frac{dz z^2}{\pi^2} \sin(\frac{x}{\Lambda\sqrt{1-z^2}}), \,\, \Lambda=\frac{L_{\Delta} L_s}{\sqrt{L_{\Delta}^2+ L_s^2}}. 
\label{J}
\end{equation}
We recall that $\Lambda$ is the characteristic spatial scale of the edge spin density. 
For $x\ll \Lambda$ we have $J(x)\propto x/ \Lambda$.  In the opposite limit $x\gg \Lambda$, we obtain $J(x)\propto (\Lambda/x)^{3/2}\cos [(x/ \Lambda) +\pi/4]$. 

\par
{\it Weak SO coupling: $L_s \gg L_{\Delta}$.}
 To contrast our results with the usual one-band Rashba case it is instructive to consider here the limit of a weak SO coupling $2\eta k_F \ll \Delta$. 
In this limit we can calculate the difference of the distribution functions entering Eq.~(\ref{spin2}) using their standard expressions at $\eta=0$ (see also SM),  i.e.,
\begin{equation}
f_{1,2}=(eE\hbar k_y/m)\tau_{tr}(p_{1,2})\partial f_0/\partial \varepsilon, 
\label{distribution}
\end{equation}
\begin{figure}
\begin{center}
\centerline{\resizebox{2.915in}{!}{\includegraphics{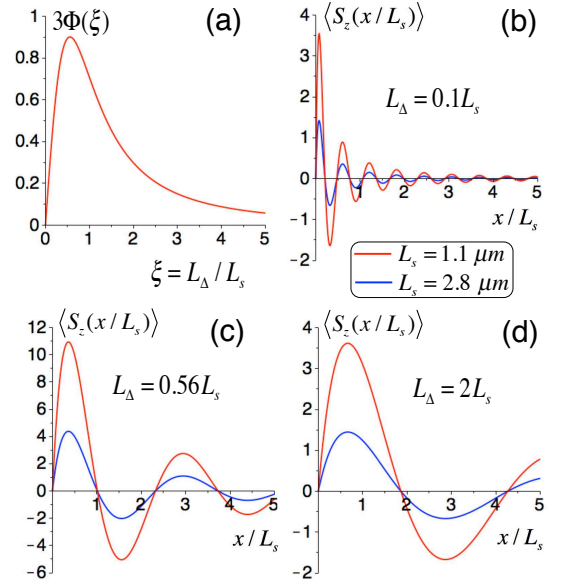}}}
\end{center}
\caption{$\Phi(\xi)$ vs. $\xi$ a) and the edge spin density $\langle S_z(x)\rangle$ in units of $10^6 cm^{-2}$ for distinct ratios $\L_{\Delta}/L_s$ and two different values of $L_s$  b)-d), as a function of $x/L_s$. Note that $\Phi(\xi)$ has a maximum at $\xi \sim 0.56 = L_{\Delta}/L_s$. 
The amplitude of the oscillations is reduced as $L_s$ increase (cf. blue and red curves in b)-d).) } 
\label{fig:Spin2}
\end{figure}
where $f_0$ is the Fermi function, the electric field $E$ is directed along the $y$-axis, and $\tau_{tr}(p)$ is the momentum-dependent  transport scattering time calculated within the Born approximation due to impurity scattering in the bulk.  The values of $p_1,p_2$ are related through $\varepsilon_{1}(p_1)=\varepsilon_{2}(p_2)=\varepsilon=\varepsilon_F$, Fig.~1(b).  
Using the condition  $k_F d\gg 1$ (small-angle scattering in bulk), we obtain $(\tau_{tr}(p_1)-\tau_{tr}(p_2))/\tau_{tr}\approx 3(p_1-p_2)/p_F \approx (3/(k_FL_{\Delta})$. \cite{note}
We note that compared to the usual Rashba one-band case  the difference of the distribution functions considered here is finite at $\eta=0$, and is of the first order in $p_1-p_2=\hbar/L_{\Delta}$.  Since the SO coupling is weak, the probability of the spin flip is small which shows up as  
the small factor $\sin^2\theta \approx L_{\Delta}^2/L_s^2 \ll 1$  in Eq.~(\ref{spin2}), and finally we obtain
\begin{equation}
\langle S_z(x) \rangle = 3 k_E \frac{L_{\Delta}}{L_s^2}J(x/\Lambda),  
\label{small}
\end{equation}
which coincides with the result which follows from Eq.(\ref{result}) in the limit $\xi \to 0$.
\par
The calculated edge spin density Eq.~(\ref{result}) is maximal at $ L_{\Delta}\approx L_s$ when it is of the order of $k_E /L_s$. 
With decreasing the gap ($L_s < L_{\Delta}$) the spectrum approaches the usual Rashba model type (doubly degenerate), and because of the specific cancellation inherent  in that model $\langle S_z(x)\rangle$  decreases in magnitude as  $k_E L_s/L_{\Delta}^2$  [see Fig.~2(a)], 
finally approaching the limit calculated in Ref. \onlinecite{Khaetskii} given by $\simeq (k_E /L_s) (d^2/L_s^2)$ [see also Eq.~(\ref{one_band})].
Thus for a given strength of the SO interaction, the magnitude of the edge spin density has non-monotonic dependence as a function of the $\L_\Delta$ (or $\Delta$),  Fig.~2(a). 
We note that  if one takes  for the ratio $d/L_s =0.1$, then the edge spin density obtained in Ref.~\cite{Khaetskii} for the 
 usual Rashba system with one sub-band equals in magnitude  the density which follows from Eq.~(\ref{result}) at $\xi \approx 35$,  where the latter is three orders of magnitude smaller than its maximal value at $\xi=0.56$. 

\par
{\it Comparison with the experiment.}
The experimental estimate of $L_{\Delta}$ is $\approx 1.4 \times 10^{-5}$ cm.   For $L_s$ we take two characteristic lengths $ 1.1\mu m $ and $2.8 \mu m $.  Note that the  corresponding values of $\eta$ are consistent with the ones obtained from the theoretical calculations \cite{jiyong-egues} for structures similar to that used in the experiment of Ref.~\cite{Hernandez}. Thus the value  $\xi=0.1$ will reasonably correspond the above chosen lengths.
 Calculating $ \tau_{tr} $ from the mobility $1.9 \times 10^{6} cm^2$/Vs , and using   $E=0.05$ mV/$\mu$m for the electric field in the quasi-ballistic region of the sample (both the mobility and $E$ are taken from Ref.~\cite{Hernandez}), we plot  $\langle S_z(x) \rangle$, see Fig.~2(b).
 The exact experimental value of the edge spin density is not known; the authors of  Ref.~\cite{Hernandez} have estimated the threshold minimal value compatible with their observation as $3\times 10^{6} cm^{-2}$.   
Hence this  number is consistent   with our  calculation.
In addition, we stress that the procedure just described, i.e.,  comparison of our theoretical predictions for the edge spin density with the experimental value of this quantity,  allows one to extract the value of $\eta$.

\par
In conclusion, using a Rashba-like SO interaction arising from the coupling between two sub-band states of opposite parities in a symmetric two-subband quantum well, 
we have explained  the great difference  between the edge spin density in a bilayer structure as compared to the one in a single-layer configuration observed in the experiment of Ref.~\cite{Hernandez}. 
The presence of the gap between the two sub-bands governs the effective strength of the inter-subband SO interaction and  changes
drastically the picture of the edge spin accumulation. 
Thus by varying the gap value 
one can easily proceed from the regime of strong spin accumulation to the regime of weak spin
accumulation.  This opens up 
the possibility for the design  of  new spintronic devices.

\par
We acknowledge financial support from FAPESP (Funda\c c\~ao de Apoio \`a Pesquisa do Estado de S\~ ao Paulo). Helpful discussions with G. Gusev and F. G. G. Hernandez are greatly appreciated. A. Khaetskii is also grateful
to Instituto de F\'\i sica de S\~ao Carlos of the University of S\~ao Paulo  for the hospitality.


\begin{thebibliography}{99}
\bibitem{note2} One-subband or two-subband configurations were realized experimentally using symmetric quantum wells  of different widths. 
A $14$ nm wide sample contained only one subband but a $45$ nm wide sample contained two (even and odd)  subbands. 
Due to the Coulomb repulsion of the electrons in the wide quantum well, the charge distribution resulted in a bilayer electron system with a soft
barrier inside the well.  Tunneling through this barrier leads to the symmetric-antisymmetric splitting $\Delta_{SAS}$ for which we use the notation $\Delta$.
\bibitem{EF} For the two-subband wide-well sample of Ref.~\cite{Hernandez} the total electron density $n=9.2\times 10^{11}$ cm$^{-2}$ and  the Fermi wave vectors $k_{F,1} = k_{F,2}=k_F=\sqrt{2\pi n/2)}=1.7\times 10^6$ cm$^{-1}$. The Fermi energy is $\varepsilon_F=\hbar^2k_F^2/2m=16.4$ meV (assuming $m=0.067m_0$ for a GaAs well). Note that $\varepsilon_F \gg \Delta=1.4$ meV in Ref.~\cite{Hernandez}.
\bibitem{note1}
A fast contribution to the edge spin density which oscillates as function of $ x$ with
$2k_F$ wave vector, gives parametrically smaller contribution to the total spin $\int_0^{\infty}dx \langle S_z(x) \rangle$, and we omit it. 
 \cite{Khaetskii}
\end{thebibliography}
\end{document}